\documentclass[review]{elsarticle}

\usepackage{lineno,hyperref}
\modulolinenumbers[5]

\journal{Journal of \LaTeX\ Templates}

\usepackage{multirow}
\usepackage{amsmath}
\DeclareMathOperator*{\argmax}{arg\,max}                       
\usepackage{graphicx,subfigure}
\usepackage{graphicx,verbatim,epstopdf,enumerate}
\usepackage{algorithm,algcompatible,amsmath}
{}
{}
{}
\algnewcommand\INPUT{\item[\textbf{Input:}]}%
\algnewcommand\OUTPUT{\item[\textbf{Output:}]}%
\usepackage{times}
\usepackage{url}
\usepackage{latexsym}
\usepackage{amsmath}
\usepackage{amssymb}
\usepackage{tabularx,ragged2e}
\usepackage{array}
\usepackage{varwidth}
\newcolumntype{Y}{>{\centering\arraybackslash}X}
\usepackage{booktabs}
\usepackage{float}
\usepackage{ifpdf}
\usepackage{balance}
\usepackage{caption}

\begin{document}

\begin{frontmatter}

\title{Multilingual and Multimode Phone Recognition System for Indian Languages}

\author{Kumud Tripathi, M. Kiran Reddy and K. Sreenivasa Rao}

\tnotetext[mytitlenote]{Fully documented templates are available in the elsarticle package on \href{http://www.ctan.org/tex-archive/macros/latex/contrib/elsarticle}{CTAN}.}

\address{Department of Computer Science and Engineering, \\Indian Institute of Technology Kharagpur, India}
\fntext[myfootnote]{Indian Institute of Technology Kharagpur, Kharagpur, West Bengal 721302, India.}

%
\cortext[mycorrespondingauthor]{Kumud Tripathi: kumudtripathi.cs@gmail.com}

\begin{abstract}
The aim of this paper is to develop a flexible framework capable of automatically recognizing phonetic units present in a speech utterance of any language spoken in any mode. In this study, we considered two modes of speech: conversation, and read modes in four Indian languages, namely, Telugu, Kannada, Odia, and Bengali. 
The proposed approach consists of two stages: (1) Automatic speech mode classification (SMC) and (2) Automatic phonetic recognition using mode-specific multilingual phone recognition system (MPRS). In this work, vocal tract and excitation souce features are considered for speech mode classification (SMC) task. SMC systems are developed using multilayer perceptron (MLP). Further, vocal tract, excitation source, and tandem features are used to build the deep neural network (DNN)-based MPRSs. The performance of the proposed approach is compared with mode-dependent MPRSs. 
Experimental results show that the proposed approach which combines both SMC and MPRS into a single system outperforms the baseline mode-dependent MPRSs.

\end{abstract}


\end{frontmatter}


\section{Introduction}\label{intro}

The objective of the phone recognition system (PRS) is to convert a speech signal into a sequence of phones. The PRS can be developed using data from single language \cite{Manjunath2015, Pradeep2016}, or multiple languages \cite{Scanzio2008, Burget2010, Manjunath2019}. A PRS trained with data from more than one language is known as multilingual PRS (MPRS). In the literature, there are very few studies on multilingual phone recognition. Schultz et al., proposed multilingual acoustic models for read speech recognition \cite{Scanzio2008, Burget2010}. Here, large vocabulary speech recognition systems are investigated for 15 languages. In \cite{Kumar2005}, a unified approach for the development of the hidden Markov model (HMM) based multilingual speech recognizer is proposed. The study considered two acoustically similar languages, Tamil and Hindi along with an acoustically very different language, American English. Here, Bhattacharyya distance measure is used to group the acoustically similar phones across the considered languages.  In \cite{Gangashetty2005}, a syllable-based multilingual speech recognizer is developed using 3 Indian languages: Telugu, Tamil, and Hindi. Here, vowel onset points are used as anchor points to derive the syllable-like units. Similar consonant-vowel units across the considered languages are merged to train a multilingual speech recognizer. Mohan et al. \cite{Mohan2014} developed a small vocabulary multilingual speech recognizer using two linguistically similar Indian languages: Marathi and Hindi. Here, the multilingual speech data collected over mobile telephones are used for training a subspace Gaussian mixture model. The speaker variations are handled by employing a cross-corpus acoustic normalization technique. In \cite{Manjunath2019}, deep neural network (DNN)-based MPRS is developed using MFCC and tandem features, for read speech. The study considered speech utterances from 4 Indian languages, namely, Telugu, Kannada, Odia, and Bengali. Here, the acoustically similar units from various languages are grouped by using the international phonetic alphabet (IPA) transcription for training the MPRS.

In general, speech can be broadly classified into two modes, namely, read and conversation modes \cite{Batliner1995, Blaauw1991, Dellwo2015}. Read mode is a formal mode of speech where a person speaks in a constrained environment, for example, news broadcasts from television. On the other hand, conversation speech is spontaneous, informal, unstructured, and unorganized.  Generally, an MPRS is trained and tested using data from the same speech mode (read or conversation), viewed as mode-dependent MPRS. The performance of mode-dependent MPRS will be affected when input utterance belongs to a different mode of speech. The reason for this is the mismatch in the acoustic characteristics of speech signals across various modes of speech. {The Indian TV news channel such as News 18, Zee News, etc. may represent the real-time scenario where read and conversation modes come together. 
The read speech is delivered by a single speaker such as newsreader and whereas the conversation speech is delivered by a group of people discussing the issues on a particular topic. Whenever the data is captured for this scenario, these two modes of speech come together and will reduce the accuracy of phone recognition task. To improve the accuracy of phone recognition system in multimode speech signals, a framework is required to decode both the modes (read, conversation) accurately.
It is also observed from a renowned video-sharing website: YouTube that the speech transcription is auto-generated for the English news channel; however, no speech transcription produced for news channel based on Indian languages. So, the current study is motivated by the recognition of speech for Indian languages in read and conversation modes.}

There exists no previous work related to MPRS, which accepts speech utterances from multiple modes at the input and generates a sequence of phonetic units at its output. In this work, we develop a framework for automatically recognizing phonetic units present in a speech utterance of any language spoken in any mode. 
The proposed approach combines the speech mode classification (SMC) system and MPRS into a single framework. The SMC system is used as a front-end to recognize the mode of the input speech utterance, which is then given as input to corresponding mode-specific MPRS to decode the phonetic units. 
Previous studies have explored the characteristics of various speech modes, such as neutral (read), soft, loud, whispered, shouted, etc. Hansen et al. \cite{Hansen_1989} have analyzed the characteristics of neutral mode and further compared it with loud, soft, and stressed mode. Rostolland has presented the phonetic structure and acoustic information of shouted mode in \cite{Rostolland1982a, Rostolland1982}, respectively. Zhang et al. \cite{Zhang2007} have studied and compared the vocal characteristics of neutral, whispered, soft, shouted, and loud speech modes. However, no previous work has explored the conversation speech mode. An utterance in conversation mode may include other speech modes such as neutral, soft, loud, whispered, shouted, etc. Hence, it's important to study the acoustic characteristics of conversation mode as well as the acoustic difference between the conversation and read modes for recognizing speech in some real-life applications. Therefore, read, and conversation modes are considered in this study for developing the speech mode classification systems.  
As multilayer perceptron (MLP) is standard classifier for various speech applications \cite{Dede2010,Vinyals2011,Koolagudi2012}, thus, the SMC systems are developed using MLP in this work. 
The vocal tract and excitation source information have been investigated for developing SMC system. While Mel-frequency cepstral coefficients (MFCCs) represent the vocal tract information, supra segmental level epoch strength contour (ESC) and pitch contour (PC) represents the excitation source information. 
Further, the scores of the vocal tract and excitation source systems are combined to improve the performance of the SMC system. The mode dependent MPRSs are developed using DNNs. The MPRSs are trained using a combination of MFCCs, tandem features, and excitation source features. In this work, the development of SMC systems and MPRSs is carried out using data from four Indian languages: Telugu, Kannada, Odia, and Bengali. 
The significance of the proposed framework is shown by comparing the performance with two mode-specific MPRSs.

The workflow of the paper is as follows. Section \ref{corpus} provides the details about the speech corpora used in this work. 
Feature extraction techniques are described in Section \ref{feat}. The significance of excitation source features for speech mode classification is provided in Section \ref{sig}. The development of MPRS has been elaborated in Section \ref{dev_mprs}. The proposed speech mode classification model is discussed in Section \ref{prop_msmc}. The evaluation of proposed SMC and MPRS have been carried out in Sections \ref{eva_msmc} and \ref{eva_mprs}, respectively. Finally, conclusion and future work of this paper are mentioned in Section \ref{con}.

\begin{figure*}[!t]
\begin{center}
\subfigure[]{\includegraphics[width=0.45\textwidth]{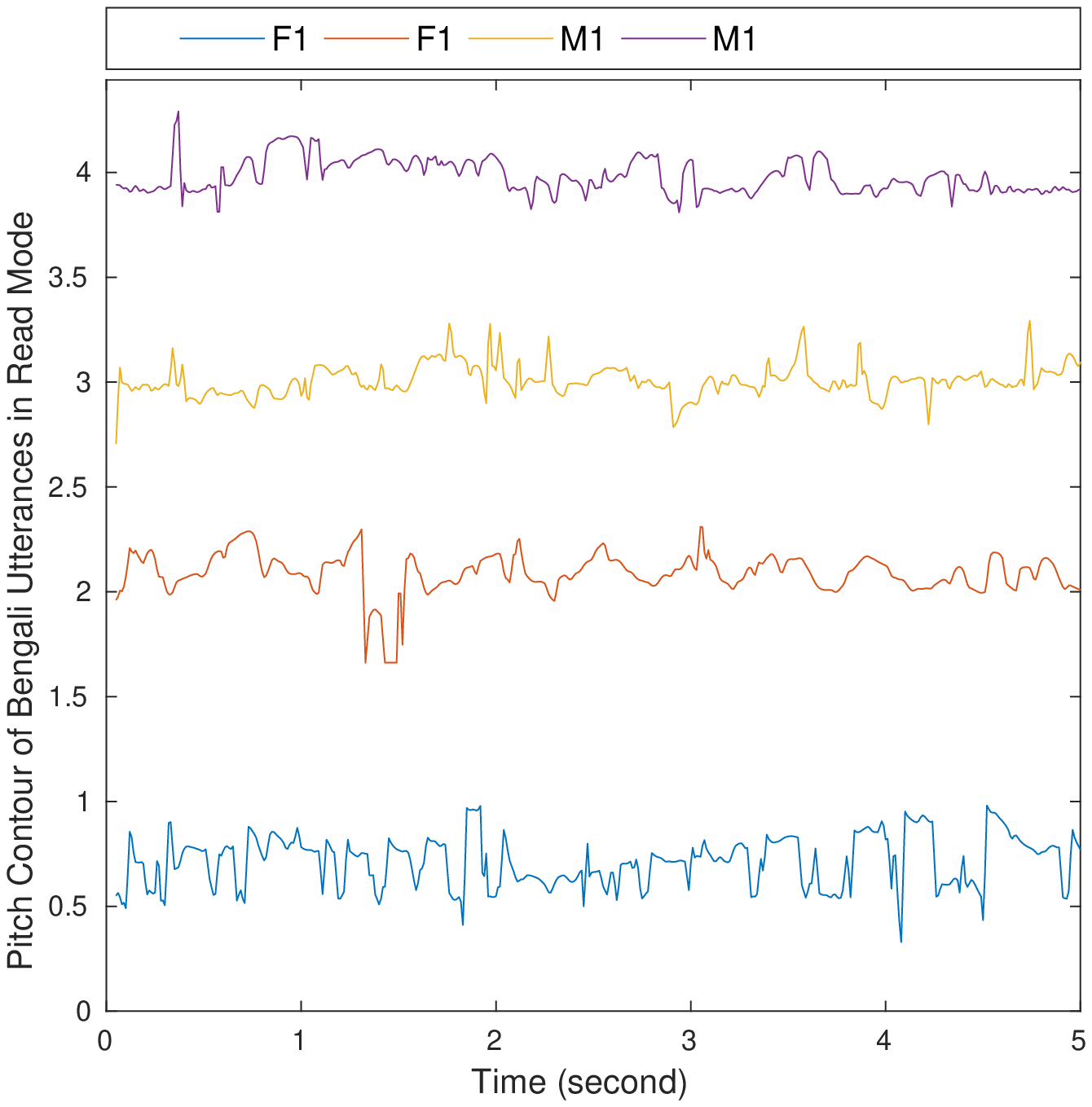}}
\subfigure[]{\includegraphics[width=0.45\textwidth]{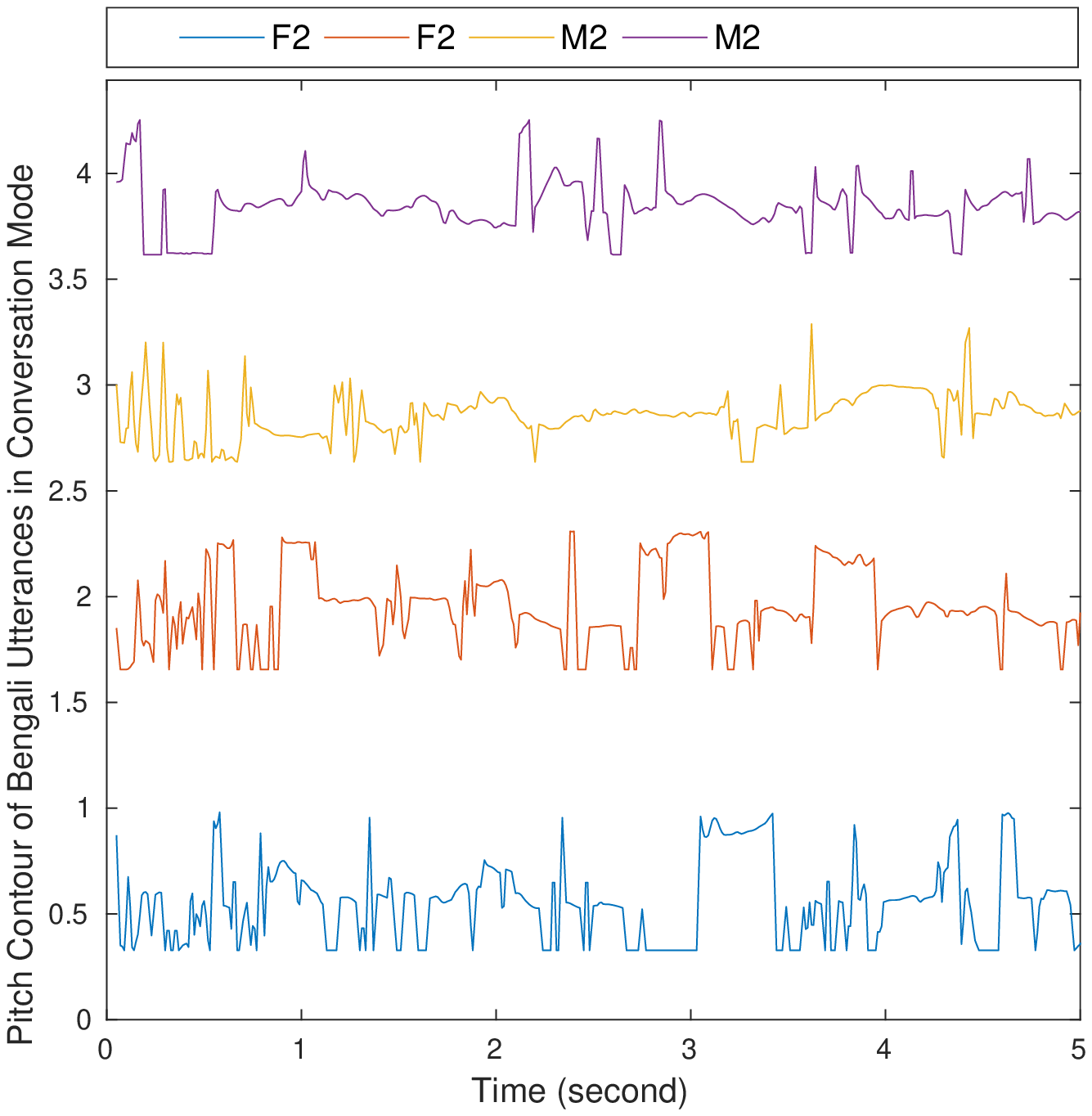}}\\
\vspace{0.5cm}
\subfigure[]{\includegraphics[width=0.45\textwidth]{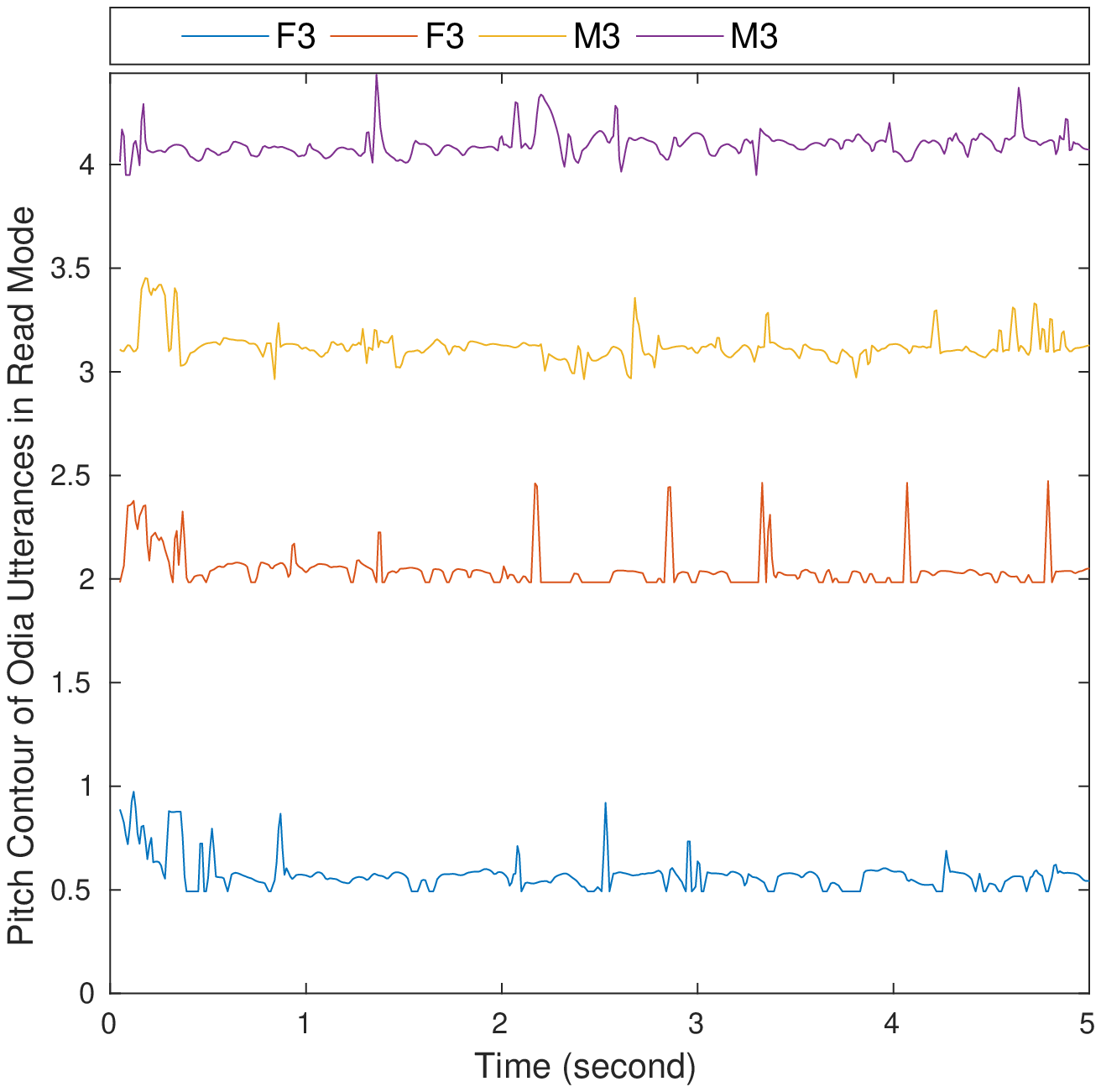}}
\subfigure[]{\includegraphics[width=0.45\textwidth]{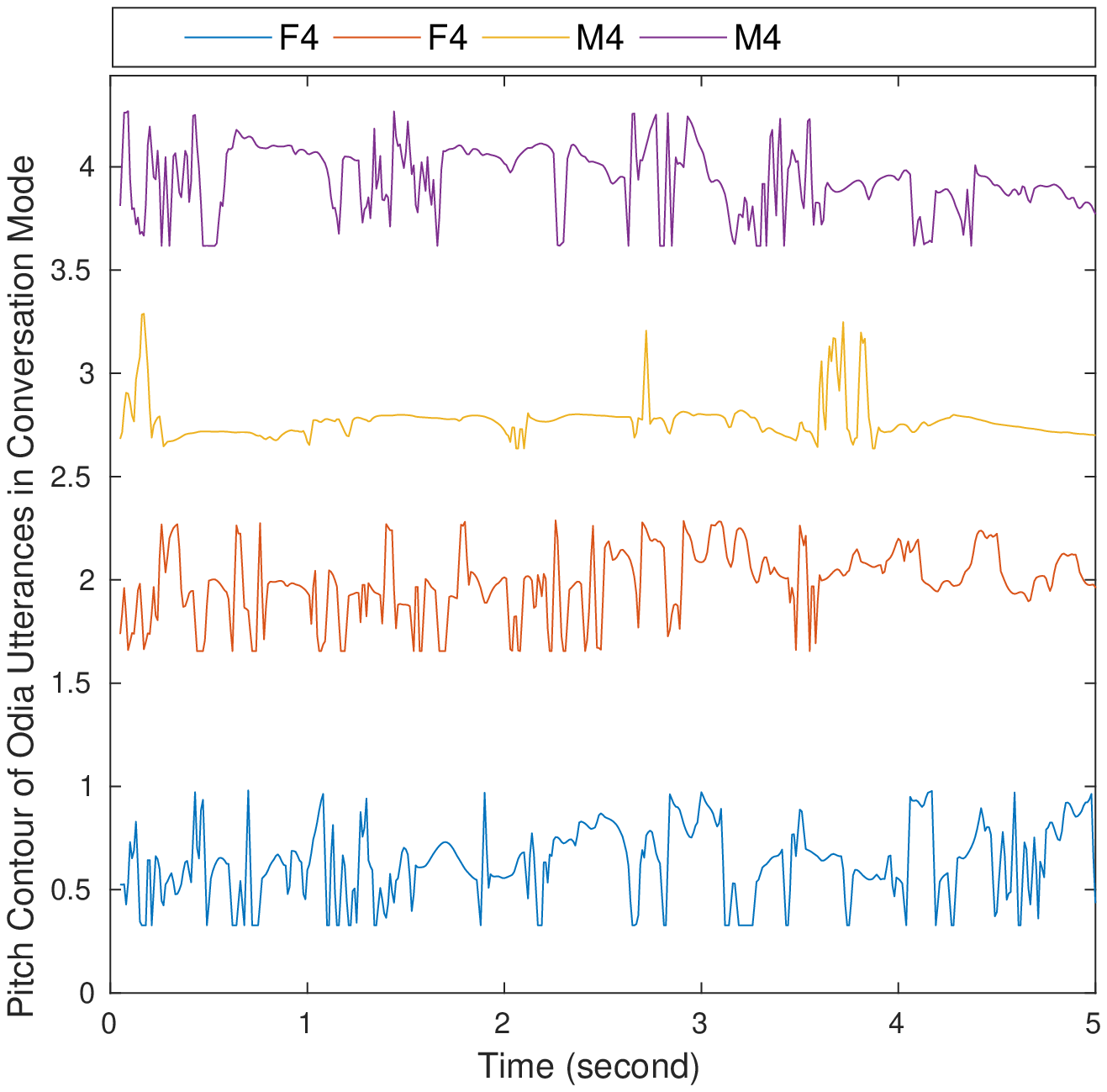}}
\caption{ Pitch contour of Bengali utterances in read and conversation modes are shown in 1(a) and 1(b), and pitch contour of Odia utterances in read and conversation modes are shown in 1(c) and 1(d), respectively. F1, F2, F3, and F4 are distinct female speakers, and M1, M2, M3, and M4 are distinct male speakers.
\label{pit}}
\end{center}
\end{figure*}

\section{Speech corpus}\label{corpus}
In this work, speech corpora of four Indian languages: Telugu, Kannada, Odia, and Bengali is considered for developing both speech mode classification models and MPRSs. The speech corpora are collected as a part of consortium project titled \textit{Prosodically guided phonetic engine for searching speech databases in Indian languages} supported by DIT, Govt. of India. The corpora contain speech data in read and conversation modes. In this study, read speech is collected from news reading, and the conversational speech is collected from {news interview}. Complete details of speech corpora are given in \cite{Kumar2013, Shridhara2013}. The speech signals are sampled at a rate of 16 KHz with 16 bits per sample. The duration of each wave file varies between 6 to 8 sec. For all the speech signals, the phonetically and prosodically rich transcription is derived using the International Phonetic Alphabet (IPA) chart. An IPA transcription provides one symbol for every unique sound unit independent of the language information. As a pre-processing step for SMC, we have removed silence regions from the speech signals, and each speech file is chopped to have a fixed duration of 5 sec. 
Table \ref{corp} details the number of male and female speakers used for training and testing. The first column contains the name of the languages, and the corresponding mode name is listed in the second column. Next two columns show the number of male and female speakers for training the models. Last two columns depict the number of male and female speakers for testing the models. 
Altogether, there are 12 distinct speakers for each mode of a language for training and 6 distinct speakers for each mode of a language for testing. Each training speaker has spoken 150 utterances whereas; each testing speaker has spoken 50 utterances. Note that speakers considered for training and testing are different as well as randomly selected.  

\begin{table}[h]
\begin{center}
\caption{The number of male and female speakers (for training and testing) from read and conversation modes of Bengali, Odia, Telugu and Kannada languages. (Abbreviation- SM: Speech Mode, M: Male, F: Female)\label{corp} }
\scalebox{1}{
\begin{tabular}{|cc|cc|cc|}
\hline
\multirow{2}{*}{\bf{Language}} & \multirow{2}{*}{\bf{SM}} & \multicolumn{2}{c|}{\bf{\# Speakers (Training)}} & \multicolumn{2}{c|}{\bf{\# Speakers (Testing)}} \\\cline{3-6}
 & & \bf{M} &\bf{F}& \bf{M} & \bf{F}\\ 
\hline
\multirow{2}{*}{Bengali (B)}& Conv & 6 & 6& 3&3 \\
& Read  & 6 & 6& 4&2 \\

\multirow{2}{*}{Odia (O)}& Conv & 5 & 7& 3&3  \\
& Read & 6 & 6& 3&3  \\
 
\multirow{2}{*}{Telugu (T)}& Conv & 6 & 6& 2&4  \\
& Read & 6 & 6& 3&3  \\

\multirow{2}{*}{Kannada (K)}& Conv & 6 & 6& 3&3  \\
& Read & 5 & 7& 3&3 \\
\hline
\end{tabular}}
\end{center}
\end{table}

\section{Feature Extraction}\label{feat}
This section describes the methods for extracting the features for speech mode classification and multilingual phone recognition systems. The vocal tract and excitation source features explored for developing the speech mode classification model are discussed in Section \ref{excit}. The vocal tract, excitation source, and tandem features for training the multilingual PRS are explained in Section \ref{mpdss}. 

\subsection{Features for Speech Mode Classification}\label{excit}

In this work, SMC models are developed using vocal tract and excitation source features. 
The 13-dimensional MFCCs along with $\Delta$ and $\Delta\Delta$  coefficients are used to capture the vocal tract information. The $\Delta$ and $\Delta\Delta$ coefficients corresponds to the first and second order derivative of MFCCs, respectively. The vocal tract parameters are extracted at frame level by considering a frame size of 25 ms and a frame shift of 10 ms using the Hamming window.  

The excitation source features describe the variations in the vibration of the vocal folds while producing voiced segments of speech. In this work, the speech signal is parametrized for 100 ms frame at the supra-segmental level to capture the mode-specific excitation source information. Pitch contour (PC) and epoch strength contour (ESC) represent the supra-segmental level excitation source information. Pitch represents the vibration frequency of the vocal cords during the speech production. Speakers utilize pitch to demonstrate the salience of words such as a higher pitch implies that the word is more vital than other words in a speech utterance. In this work, the pitch and epoch strength contours are calculated using zero frequency filtering (ZFF) approach \cite{Murty2008}. The ZFF method tries to estimate the pitch and epoch strength by finding epoch locations in the speech.
The interval between two successive epochs represents the pitch period ($t_0$). The reciprocal of pitch period ($t_0$) will give the pitch ($p_0=\frac{1}{t_0}$) \cite{Yegnanarayana2009}. The epoch strength corresponds to the slope around the positive zero crossings corresponding to the epoch locations present in the ZFF signal. The slope is estimated as the difference between the successive samples of ZFF signal around the zero crossings \cite{Murty2008}.

The vocal tract and source features are extracted at the frame level. The processing of excitation source features at frame level may get similar variation in considered speech modes. {However, its pattern at the sentence level is almost different among the speech modes, and also not dependent on the speaker, for mode classification.} This justification can be validated by visualizing the pitch patterns given in Figure \ref{pit}. Hence, it's significant to process excitation source details at the sentence level and vocal tract details at the frame level. Therefore, in this work, mode discriminative characteristics of the vocal tract and source features are studied separately. The model developed using MFCCs are trained and tested at the frame level. Afterward, majority voting \cite{Lam1997} is applied for taking decision at sentence level. In the excitation source model, training and testing are done at sentence level. Further, the scores generated at sentence level from the vocal tract and excitation source models are combined using a weighted score fusion technique for enhancing the total performance of the SMC model.

\subsection{Features for Multilingual Phone Recognition System}\label{mpdss}
 In this work, the multilingual phone recognition system is trained using the vocal tract, excitation source, and tandem features. To capture the vocal tract information for MPRS, a 13-dimensional MFCCs along with $\Delta$ and $\Delta\Delta$ coefficients are extracted at a frame size of 25 ms with an overlap of 10 ms.

\begin{table}[!htbp]
\begin{center}
\caption{Correlation coefficients across the speech modes (SM) of multiple languages using pitch contour (PC) and epoch strength contour (ESC).\label{corr} }
\scalebox{1}{
\begin{tabular}{|c|cc|cc|cc|}
\hline
\multirow{2}{*}{\bf{ID}}&\multirow{2}{*}{\bf{Language}} & \multirow{2}{*}{\bf{SM}} & \multicolumn{2}{c|}{\bf{PC}} & \multicolumn{2}{c|}{\bf{ESC}}\\\cline{4-7}
& & & \textbf{WM} &\textbf{BM}& \textbf{WM} &\textbf{BM}\\ 
\hline
\multirow{2}{*}{1}&\multirow{2}{*}{Bengali (B)}& Conv & 0.37 & 0.14& 0.63&0.06 \\
&& Read & 0.53 & 0.14& 0.67&0.06 \\

\multirow{2}{*}{3}&\multirow{2}{*}{Odia (O)}& Conv & 0.42 & 0.15& 0.64&0.02\\
&& Read & 0.56 & 0.15& 0.70&0.02 \\

\multirow{2}{*}{2}&\multirow{2}{*}{Telugu (T)}& Conv & 0.32 & 0.12& 0.52&0.04 \\
&& Read & 0.44 & 0.12& 0.68&0.04 \\


\multirow{2}{*}{4}&\multirow{2}{*}{Kannada (K)}& Conv & 0.21 & 0.13& 0.47&0.07 \\
&& Read & 0.26 & 0.13& 0.61&0.07 \\

\multirow{2}{*}{5}&\multirow{2}{*}{K-T-B-O}& Conv & 0.21 & 0.12& 0.59&0.03 \\
&& Read & 0.35 & 0.12& 0.67&0.03 \\
\hline
\end{tabular}}
\end{center}
\end{table}
 The excitation source features are extracted at the segmental level. Here, Mel power differences of spectrum in sub-bands (MPDSS) and residual Mel-frequency cepstral coefficients (RMFCCs) are used as source features. The steps for deriving MPDSS feature are as follows. First, compute the LP residual of the speech signal. Further, the LP residual signal is segmented into frames of 25 ms with 10 ms overlap. For every LP residual frame, LP residual spectrum is computed by using a discrete Fourier transform (DFT). 
The power spectrum is computed as the square of the magnitude of the spectrum of the LP residual signal. The power spectrum is then processed through $m$ Mel-filter bank. For each filter bank, PDSS coefficients are computed, and all extracted PDSS coefficients represent MPDSS coefficients. In this study, 25-dimensional MPDSS coefficients are chosen for training the MPRSs.

Residual Mel-frequency cepstral coefficients (RMFCCs) are
generated when a residual signal is utilized for deriving MFCC
features. The framewise RMFCCs are computed using 10th order LP residual. This process is similar to the estimation of speech MFCCs except that the input is residual signal instead of the speech signal.
In this work, 13-dimensional RMFCCs along with $\Delta$ and $\Delta\Delta$ coefficients are computed using the frame size of 25 ms with 10 ms frame shift for developing MPRSs.

Tandem features are the phone posteriors generated
after training a classifier using spectral features. Tandem features also known as discriminative features as they are produced from a discriminative classification model. In this work, we have followed the same procedure as discussed in \cite{Manjunath2019} for extracting the tandem features. The spectral feature MFCCs are used for training the discriminative classifier deep neural network (DNN) for extracting the phone posteriors.
The dimension of the feature vector is 44, which represent the total number of unique phones present in the considered speech corpus (discussed in Section \ref{corpus}).

\section{Significance of excitation source features for mode classification}\label{sig}
The importance of pitch and epoch strength contours for mode classification task is presented by their respective correlation coefficients (CCs) \cite{Benesty2009} for within and between modes across the languages, in Table \ref{corr}. Correlation coefficients calculate the strength of relationships among the speech signals. Assume $R$, and $Q$ are two finite duration speech signals. The CC between the $R$ and $Q$ can be calculated as follows:
 \begin{equation}
C(R,Q)=\frac{1}{L}\sum_{i=1}^{L}\Bigg(\frac{\overline{R_i-\mu_R}}{\sigma_R}\Bigg)\Bigg(\frac{\overline{Q_i-\mu_Q}}{\sigma_Q}\Bigg)
\end{equation}
Where $\mu_R$ and $\sigma_R$ are mean and standard deviation of $R$, respectively, and $\mu_Q$ and $\sigma_Q$ are mean and standard deviation of $Q$. $L$ is the number of samples in each signal.
The value of correlation coefficient is maximum when signals are similar and minimum when signals are orthogonal. If the signals have no relation, then the value of the coefficient is $0$. 

For each language, four speakers data (two speaker's data
per mode) is considered for computing the correlation coefficients. Note that the speakers are distinct for each language and each speaker has uttered 50 distinct utterances.
 To normalize the speaker variability across languages, the mean subtraction is imposed on the feature vectors across all languages. 
The correlation coefficients among modes of monolingual speech signals are depicted using ID: 1,2,3,4, whereas among modes of multilingual speech signals are shown with ID: 5. The values in the fourth and fifth columns of the Table \ref{corr} denote the CCs within the modes (WM) and between the modes (BM) using pitch contour. Similarly, the values in the sixth and seventh columns of the Table \ref{corr} denote the CCs within the modes (WM) and between the modes (BM) using epoch strength contour. 
For computing the average CC within the mode of a language, 50 distinct utterances spoken by speaker-1 of a mode is correlated with 50 distinct utterances spoken by speaker-2 of a mode. However, the average CC between the mode is computed using 50 distinct utterances from each mode of a language.
In ID: 1 of Table \ref{corr}, the first value of the fourth and sixth columns presents the average CC of within conversation (abbreviated as conv) mode of Bengali language using PC and ESC, respectively. However, the first value of the fifth and seventh columns represents the average CC of conversation mode with respect to read mode of Bengali language using PC and ESC, respectively. 
A low average CC value indicates high dissimilarity and a high average CC value indicates high similarity between pitch or epoch strength contours. From the table, it can be seen that the average CC values within a mode of any language are much higher compared to the CC values with respect to other modes. This depicts that the pitch and epoch strength contours have significant mode discrimination capability.

In Table \ref{corr}, ID: 5 (K-T-B-O) shows the correlation coefficients within and between modes, across the languages. In ID: 5, the last value of the fourth and sixth columns represents the average CC within the read mode of multilingual speech using PC and ESC, respectively. However, the last value of the fifth and seventh columns presents the average CC of read mode with respect to conversation mode of multilingual speech using PC and ESC, respectively. The average CC value of read mode with respect to conversation mode (0.12) is less as compared to the CC value
of the same mode (0.35) using PC. Similar trends can be observed for conversation mode. Hence, both the PCs and ESCs are highly similar within modes and highly dissimilar between modes, across
the languages. {This analysis confirms that the pitch and epoch strength contours contain mode discriminating information.}

\section{Development of Multilingual Phone Recognition System}\label{dev_mprs}
 
A PRS trained with data from multiple languages is called as Multilingual PRS. The open-source speech recognition Kaldi toolkit (Povey et al. 2011) \cite{Povey2011} is used for building the speech recognizers. The important components in building the phone recognition system using Kaldi toolkit are feature extraction, acoustic and language modeling, and decoding of phone sequence. The vocal tract, excitation source, and tandem features (discussed in Section \ref{mpdss}) are used for developing the multilingual phone recognition systems. DNNs are trained as in Zhang et al. \cite{Zhang2014} for acoustic modeling. For language modeling (LM), the prior probability of a phone sequence (also known as LM score) is estimated by learning the relation between phones from the training data. The language model will generate a more accurate score, in case of having prior information about the speech task \cite{Yu2015}. In this work, the effect of LM is also analyzed on the performance of the MPRS. The speech corpus described in Section \ref{corpus} is considered for training MPRSs.

The objective of an MPRS is to find the sequence of phones $\hat{H}$, whose likelihood to a given sequence of feature vectors $A$ is maximum. The Eq. 2 shows  the mathematical formulation for estimating the decoded phone sequence:

\begin{equation}
\hat{H}=\argmax_H P(H|A)
\end{equation}
 The term $P(H|A)$ can be interpreted by applying Bayes' rule as, 
\begin{equation}
P(H|A)=\frac{P(A|H)P(H)}{P(A)}
\end{equation} 
 
The acoustic modeling is used for calculating the $P(A/H)$ likelihood, whereas $P(H)$ the prior probability of phone sequence is computed using language modeling. $P(A)$ shows the prior probability of the feature vector. It can be avoided in the equation (2) as it is independent of the acoustic and linguistic results. In the current work, the bi-gram language model \cite{Brown1992} is used for determining the $P(H)$. The final decoding sequence can be computed as follows:
\begin{equation}
\hat{H}=log P(A|H)+\alpha logP(H)
\end{equation}
Where $\alpha$ is the LM scaling factor which is applied to balance between the acoustic and language model scores.

\subsection{Baseline MPRS}\label{baseline_mprs}

The block diagram of the baseline MPRS is shown in Figure \ref{prs}. 
The MPRS explored in \cite{Manjunath2019}, is considered as our baseline MPRS which includes MFCC and tandem features for developing MPRS using four Indian languages, namely, Telugu, Kannada, Odia, and Bengali. In \cite{Manjunath2019}, authors have developed MPRS for read mode speech, {we referred it as} read-baseline MPRS. In this work, we have developed separate MPRS for conversation mode and named it as conversation-baseline MPRS. The conversation-baseline and read-baseline MPRSs are developed in the same manner (as discussed in \cite{Manjunath2019}).
 The read-baseline MPRS is trained using read mode, and conversation-baseline MPRS is trained using conversation mode of speech from Telugu, Kannada, Odia, and Bengali languages.
 
\begin{table}[h]
\begin{center}
\caption{Phone recognition accuracy (\%) of read-baseline MPRS, and conversation-baseline MPRS for read and conversation modes of speech.\label{read_mprs} }
\scalebox{1.0}{
\begin{tabular}{|c|cc|}
\hline
\multirow{2}{*}{\bf{Model}}  & \multicolumn{2}{c|}{\bf{Accuracy (\%)}} \\ \cline{2-3}
  & \bf{Read} & \bf{Conversation}\\ 
\hline

{Read-Baseline MPRS}& 64.45 &33.50 \\
{Conv-Baseline MPRS}& 30.30 &64.23 \\
\hline
\end{tabular}}
\end{center}
\end{table}

Table \ref{read_mprs}, shows the recognition accuracies of read-baseline MPRS, and conversation-baseline MPRS for both read and conversation modes of speech. The read-baseline and conversation-baseline MPRSs are mode dependent systems. 
The baseline MPRSs are trained using mode dependent data from the four languages: Telugu, Kannada, Odia, and Bengali as discussed in Section \ref{corpus}. For evaluating the performance of these systems, the test data is considered from both the modes of given languages in Section \ref{corpus}. 
In Table \ref{read_mprs}, we can see that read-baseline MPRS is giving better accuracy when tested with read mode (64.45\%) as compared to conversation mode (33.50\%). Similarly, conversation-baseline MPRS is giving better performance when tested with conversation mode (64.23\%) than the read mode (30.30\%). This indicates that the performance of mode-specific MPRSs degrades when there is a mismatch in the modes of train and test utterances. As we have seen in introduction Section \ref{intro} that the speech from Indian news TV channels  may contain read and conversation mode simultaneously. In such realistic cases, the baseline systems will fail in successfully recognizing speech from different modes.
Hence, for achieving optimal performance, we need to process the speech signal associated with a particular mode through the corresponding mode-specific MPRS. One way to achieve this is by manually tagging the mode of each input utterance. However, this is complex and not feasible in a real-time scenario. Hence, it is required to develop a system which can automatically identify the mode of the input speech. This can be more advantageous if such a system is language independent. This necessity motivated us to develop a speech mode classification (SMC) system that can automatically detect the mode of input speech of any language. 
{It is noticed from the literature survey that there is no work related to multilingual speech mode classification. }
{In further section, we will discuss the proposed approach which integrates the SMC system and MPRSs into a single framework.

\begin{figure}[!h]
\centering
\includegraphics[width=\textwidth]{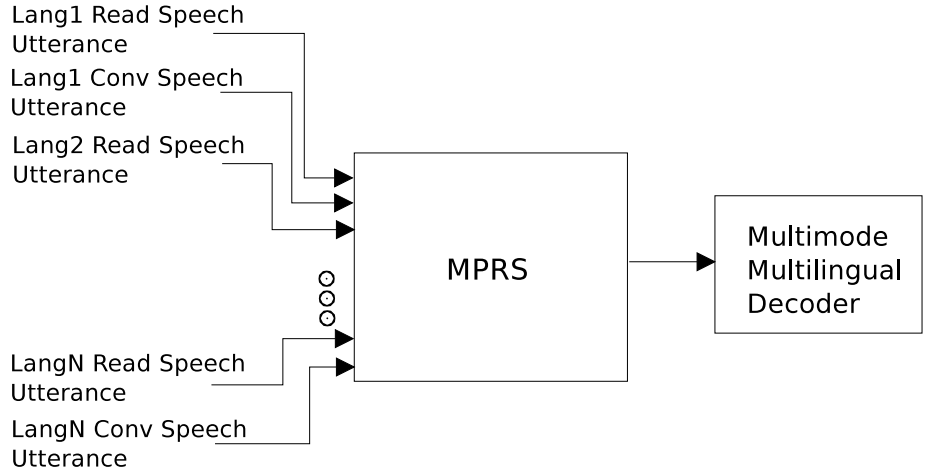}
\caption{Illustration of baseline MPRS from the multimode speech signals of multiple languages.\label{prs}}
\end{figure}

\subsection{Proposed Framework}\label{proposed_mprs}
   The block diagram of the proposed two-stage phone recognition system is shown in Figure \ref{mprs}. The first stage consists of speech mode classification, and the second stage consists of phone recognition. At the first stage, speech mode classifier is automatically detecting the mode of input test speech, i.e., read mode or conversation mode. At the second stage, phonetic units are decoded by processing the input speech signal through corresponding mode dependent MPRS. The Mode dependent MPRS is trained using phonetic data of a particular mode from four languages: Telugu, Kannada, Odia, and Bengali (discussed in Section \ref{corpus}). The proposed MPRSs are developed using the same experimental setup as in the baseline system \cite{Manjunath2019}. The difference between the baseline and proposed MPRSs is that the baseline MPRSs are developed using a combination of MFCCs and tandem features, whereas the proposed MPRSs are developed using a combination of MFCCs, tandem features, and excitation source features. The proposed model has the capability to automatically detect the mode and respective phonetic transcription of the speech signal.  This proposed framework is named as COMB-MPRS for the rest of the paper.

\begin{figure}[!h]
\centering
\includegraphics[width=1\textwidth]{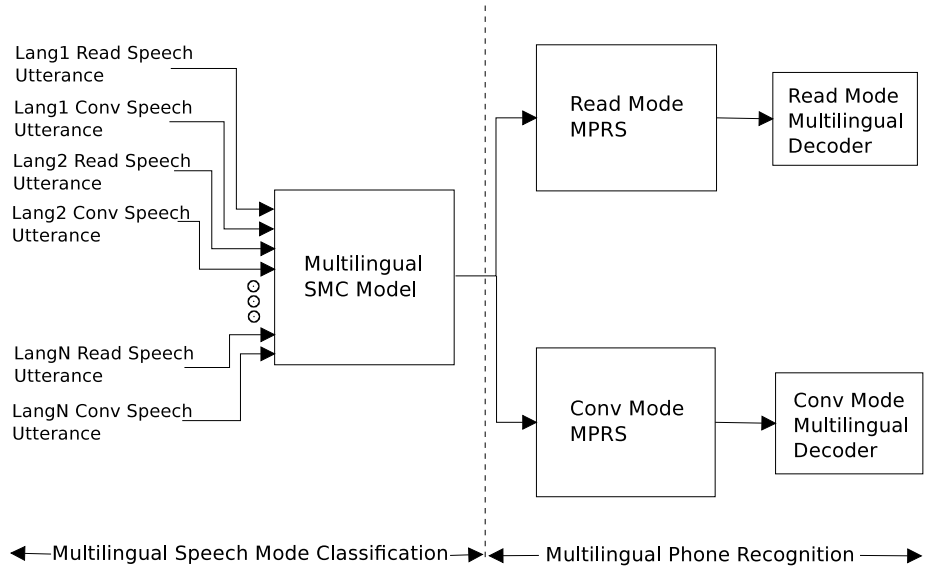}
\caption{Block diagram of the proposed framework of 2-stage MPRS.\label{mprs}}
\end{figure}

\section{Development of Proposed Multilingual Speech Mode Classification Model}\label{prop_msmc}
  {The objective of multilingual speech mode classification (SMC) model is to identify either read mode or conversation mode of speech which is spoken in any Indian language. Multilingual SMC model is a language-independent SMC model. For developing an SMC model independent of a particular language, need to include training data from that language. The multilingual SMC model is biased towards the languages used for training the model because the speech mode characteristics vary differently among the languages.
In this study, the SMC model is developed using the vocal tract and excitation source features.} As we can see in Figure \ref{msmc}, SMC models are designed at three stages. Multilayer perceptron (MLP) is used for building the classification model. 
The main reason for using MLP for building the SMC models is that it can capture the nonlinear relations hidden in the multi-dimensional feature patterns of the input utterance. MLP can also deal with the unseen input vector, significantly well.
For improving the performance of the mode classification, scores are combined at different stages, as shown in Figure \ref{msmc}. Score level fusion is described in Section \ref{score} and Sections \ref{s1}-\ref{s3} include brief introduction of each stage. 

\begin{figure*}[t]
\centering
\includegraphics[width=\textwidth]{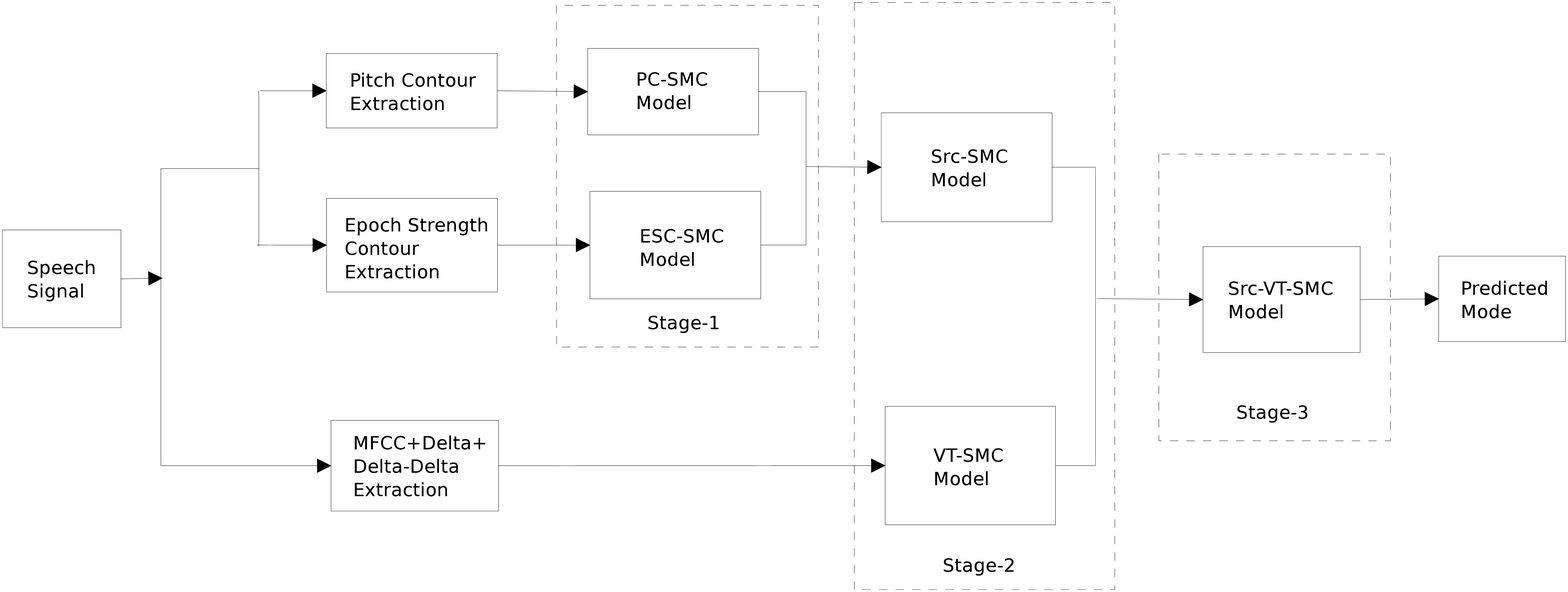}
\caption{Block diagram of SMC model using vocal tract and excitation source features.\label{msmc}}
\end{figure*}


\subsection{Multilayer Perceptron}\label{mlp}
In this work, MLP is explored for mapping a given speech signal into a speech mode by calculating invariant and discriminant features using its nonlinear processing.
MLP is a feed-forward neural network \cite{Svozil1997} with one or more hidden layers between its input and output layer. 
There are various parameters to tune in MLP, such as the number of hidden layers, the number of nodes in each layer, and learning rate. Finding these parameters is a necessary task for optimizing the MLP. 
According to \cite{Csaji2001}, an MLP with a single hidden layer can approximate any nonlinear function with optimal accuracy. However, the required number of neurons in the hidden layer is not stated in \cite{Csaji2001}. Hence, in this work, we have explored a  three-layer multilayer perceptron. The three-layered MLP is initialized with 0.005 learning rate. The \textit{tanh} non-linear function and the softmax activation function are used at the hidden and output layers, respectively. 
In this work, we have explored various network structures for both excitation source and vocal tract based speech mode classification tasks. 
For each model, the number of nodes at the input layer $p$ is equal to the dimension of input feature vectors. The number of nodes at the output layer is equal to the number of classes in a classification problem. The number of nodes at the output layer $r=2$ will be the same for all models because this is a two class problem.
The number of units in the hidden layer $q$ for each model is decided after performing experiments on huge training datasets.
The explored network structure for excitation source and vocal tract based SMC models are specified in stage-1 and 2 (see Section \ref{s1} and \ref{s2}), respectively.  
%
Further, Standard back-propagation algorithm (stochastic gradient descent) \cite{Bottou2010} is used for training the MLP to minimize the root mean squared error between the actual and the predicted outputs. 
The classification accuracy of the speech modes is analyzed by calculating the objective measure, such as root mean square error. 
%
%

\subsection{Development of SMC models at Stage-1}\label{s1}
The goal of stage-1 is to build separate SMC model for excitation source features. The excitation source features are captured from pitch and epoch strength contours of speech signals. Each feature contains distinct mode-related information. Therefore, two different SMC models are developed using MLP at stage-1: (i) SMC using pitch contour is denoted as PC-SMC model, and (ii) SMC using epoch strength contour is represented as ESC-SMC model.
The number of nodes at the input and hidden layers for developing the SMC models using the pitch contour are $p_1=500$ and $q_{1}=56$, respectively. The similar, network structure is followed for epoch strength contour based SMC model.
These two MLP models are trained and tested at sentence level.
The training process of PC-SMC and ESC-SMC models is terminated after 200 epochs as there is no substantial decrement in the error when the number of epochs is further increased.

\subsection{Development of SMC models at Stage-2}\label{s2}
  The target of stage-2 is to design the SMC model using excitation source features and examine the mode discrimination details present in the excitation source and vocal tract features. Hence, at this stage, two separate SMC models are developed using excitation source and vocal tract features. The excitation source based SMC model is developed by fusing the scores of pitch contour and epoch strength contour based SMC models from stage-1 and called an Integrated Src-SMC model. The fusion of scores is performed on test data. 
%
 Further, the vocal tract based SMC model is designed using the MFCC+$\Delta$+$\Delta\Delta$ feature named as VT-SMC model. In the VT-SMC model, the 39-dimensional feature vector is extracted for each frame of an utterance. The MLP model developed
using vocal tract features is trained and tested at the frame level. After testing, majority voting is applied for generating scores at the sentence level. The number of epochs required for training the VT-SMC model is 600. The number of nodes at input, and hidden layers are $p_{2}=39$, and $q_{2}= 21$, respectively.

\subsection{Development of SMC models at Stage-3}\label{s3}
 The objective of stage-3 is to design the single integrated SMC model. Here, the integrated model is developed by combining the scores of the excitation source and vocal tract based SMC models of stage-2 and denoted as Src-VT-SMC model. 
It is worth noting that the integrated model (Src-VT-SMC model) outperform the SMC models developed using stand-alone features at almost the same computation time.

\subsection{Score Level Fusion}\label{score}
 The score is an estimate of the relationship between the feature vectors of training and testing utterances. The score is represented in terms of the posterior probability corresponding to each class for a speech frame. Evidence generated from classifiers is dependent on the initial features. In order to reduce the significance of less efficient features and to increase the impact of more significant features, the scores can be weighted. This is known as weighted score level fusion. Adaptive weighted combination scheme \cite{Reddy2013} has been used for combining the scores of speech mode classification (SMC) models.
In this work, weights are derived by seeking the range of weights for each model present in a respective stage (see Figure \ref{msmc}) on the development set and the final optimal weights are used for evaluation. 
The idea behind the weighted score fusion is to reduce the total error on the evaluation set. The fused score is calculated as follows:

\begin{equation}
\hat{S}{(x)}=\Sigma_{i=1}^{c}w_iS_i(x)
\end{equation}

Where, i=\{1,2,..,c\} represents the number of models $c$, for development set $x$ score generated from $i$th model is shown as $S_i(x)$. Whereas, $\hat{S}{(x)}$ indicates the weighted fusion score. Weight of the $i$th model is represented as $w_i$. 
Following conditions are applied for capturing the suitable set of weights:

\begin{equation}
\Sigma_{i=1}^{c}w_i=1 \quad \textrm{and} \quad w_i \geq 0
\end{equation}

The weighting factor $w_i$ is computed at the step size of 0.01. 
For the Bengali language, we have experimented with 98 distinct sets of weighting factors for two models explored at stage 1. Similarly, 98 distinct sets of weighting factors are explored for two models present at stage 2.
 Out of 98 diverse weighting values, the considered one are those on which the best average accuracy is accounted for the integrated models.  At stage-2, best average performance of integrated Src-SMC model is obtained at weighting values of $w_1 = 0.45$ and $w_2 = 0.55$ for pitch contour and epoch strength contour based SMC models, respectively. However, at stage-3, best average accuracy of integrated Src-VT-SMC model is obtained at weighting values of $w_3 = 0.35$ and $w_4 = 0.65$ for excitation source and vocal tract based SMC models, respectively. Afterward, similar experiments are performed for each language for obtaining the best weighting values. From the experiments, it is observed that the computed weighting values of $w_1 = 0.45$, $w_2 = 0.55$, $w_3 = 0.35$ and $w_4 = 0.65$ are giving best performance for each monolingual and multilingual SMC models. Therefore, we can report that the computed weighting values are language independent and can be used as the universal set of weights for developing the SMC models.

\section{Evaluation of SMC models}\label{eva_msmc}
 In this work, the significance of excitation source and vocal tract features are explored for developing the speech mode classification models using multilayer perceptron. Here, we considered read and conversation modes of speech from four Indian languages: Telugu, Kannada, Odia, and Bengali. In this work, we have explored monolingual and multilingual speech corpora for designing the SMC models. The model trained using speech signals from two different speech modes of individual languages are named as a monolingual-SMC model. Likewise, the model developed using speech signals from two modes of four Indian languages are labelled as multilingual-SMC (MSMC) model.  
 The models developed using four monolingual speech data are treated as a baseline model for our proposed MSMC model (denoted as K-T-B-O). The framework shown in Figure \ref{msmc} is used for developing the monolingual and multilingual SMC models. 
In Table \ref{msmc_per}, we have shown the average classification performance of SMC models developed at stage-1, stage-2, and stage-3 using monolingual and multilingual speech corpora. In the first column, classification accuracies of monolingual SMCs are depicted with IDs: (1-4), and the classification accuracies of the MSMC model for individual languages are shown in ID: (5-8).  The average performance of the MSMC model across languages is shown in ID: 9. The second and third columns represents the various languages involved in training and testing the models. The columns 4-10, contains the classification accuracy for stage-1, 2 and 3 SMC models.
The detailed explanation about the performance of SMC models at each stage is given in Sections \ref{s11}-\ref{s13}. 

\begin{table*}[t]
\begin{center}
\caption{ Performance of stage-1, stage-2, and stage-3 SMC models developed using monolingual and multilingual speech corpus. \label{msmc_per} }
\scalebox{0.7}{
\begin{tabular}{|c|ll|cc|ccc|cc|}
\hline
&&\multicolumn{8}{c|}{\bf{Average Classification Performance (\%) }} \\ \hline
\multirow{2}{*}{\bf{ID}}&\multicolumn{2}{c|}{\bf{Language}} & \multicolumn{2}{c|}{\bf{Stage-1}} & \multicolumn{3}{c|}{\bf{Stage-2}} & \multicolumn{2}{c|}{\bf{Stage-3}} \\ \cline{2-10}
&Training&Testing& {PC-} & {ESC-} &{Src-}& {MFCC-} & {MFCC+$\Delta$} & {Src-SMC+} & {Src-SMC+MFCC} \\ 
& &&SMC&SMC&SMC& SMC & {+$\Delta\Delta$-SMC} &{MFCC-SMC} & {+$\Delta$+$\Delta\Delta$-SMC} \\ \hline
{1}&{Bengali (B)}&{Bengali}& 51.12 & 64.37 & 72.24&83.05& 88.67& 91.12 &93.16 \\

{2}&{Odia (O)}&{Odia}& 52.12 & 66.42 & 75.15&84.25& 89.64&92.02&94.29 \\

{3}&{Telugu (T)}&{Telugu}& 46.45 & 64.44 & 72.21&79.32& 84.78& 88.14 & 91.67 \\

{4}&{Kannada (K)}&{Kannada}& 44.19 & 60.21 & 69.43&76.72& 82.61&87.07&90.87 \\\hline

{5}&\multirow{5}{*}{K-T-B-O}&{Bengali}& 51.06 &61.45 & 70.36 & 79.63&85.12 & 89.62 & 91.95 \\

{6}&&{Odia}& 51.23 & 63.27 & 71.52 & 80.24 & 86.35 &     90.05 & 92.59 \\

{7}&&{Telugu}& 45.17 & 63.34 & 69.12 & 75.19 & 79.43 & 86.42 & 90.54 \\

{8}&&{Kannada}& 42.23 & 58.56 & 66.75 & 73.56 & 78.51 & 85.19 & 89.43 \\ 

{9}&&{\textbf{Average}}& 47.42 & 61.65 & 69.43 & 77.15 & 82.35 & 87.82 & \textbf{91.10} \\
\hline
\end{tabular}}
\end{center}
\end{table*}

\subsection{Evaluation of SMC models at Stage-1}\label{s11}
At the stage-1, two different models are developed for each monolingual and multilingual speech corpora using two excitation source features at the suprasegmental level. 
 The fourth and fifth columns of the Table \ref{msmc_per} shows the mode classification performance achieved by using suprasegmental features: pitch contour (PC) and epoch strength contour (ESC). From the results, we can see that the average classification accuracies obtained with the epoch strength contour are better than those achieved with the pitch contour for each baseline and proposed multilingual models. The similar observation is also noted while analyzing the correlation coefficients in view of mode discrimination. This represents that the epoch strength contour has better mode discrimination power than the pitch contour.

\subsection{Evaluation of SMC models at Stage-2}\label{s12}
The aim of stage-2 is to develop two different speech mode classification models by processing the full excitation source and vocal tract features. The full excitation source based model at stage-2 is designed by fusing the scores from pitch and epoch strength contour based models and termed as Src-SMC. The vocal tract based model at stage-2 are designed by using the 39-dimensional MFCC+$\Delta$+$\Delta\Delta$ feature. For analysis purpose, 13-dimensional MFCC feature is also considered for developing the SMC model.
 In columns 7 and 8, the average classification accuracies achieved with the MFCC+$\Delta$+$\Delta\Delta$-SMC model are better than those obtained with the MFCC-SMC model for each monolingual and multilingual speech corpora. Therefore, we have considered only MFCC+$\Delta$+$\Delta\Delta$ based vocal tract feature for developing the SMC model at stage-2 in Figure \ref{msmc}. Further, the average classification performance of the SMC model developed using full excitation source features is less than the models developed using MFCC+$\Delta$+$\Delta\Delta$. This indicates that the vocal tract features contain significantly higher mode specific information than excitation source.

\subsection{Evaluation of SMC models at Stage-3}\label{s13}
At this stage, two integrated speech mode classification models are developed by concatenating scores from (i) Src-SMC and MFCC-SMC models (labelled as Int-1-SMC) and (ii) Src-SMC and MFCC+$\Delta$+$\Delta\Delta$-SMC models (labelled as Int-2-SMC).
%
%
%
 In Table \ref{msmc_per}, the average mode classification accuracies determined from the integrated models are displayed in the ninth and tenth columns. The performance of integrated models is better compared to the models based on the individual feature. For example, the accuracy of the Int-2 SMC model is better than the accuracies of the Src-SMC and MFCC+$\Delta$+$\Delta\Delta$-SMC models. Among the two integrated models, the performance of Int-2-SMC model is better than that of Int-1-SMC model. Please note that the integrated models are developed using score fusion instead of feature fusion.

From the results of the Table \ref{msmc_per}, it is observed that the developed multilingual SMC models (see ID: 5-8) are providing similar results as compared to the monolingual SMC models (see ID: 1-4). Thus, a single MSMC model can replace multiple monolingual SMC models for mode classification of the speech samples belongs to any language. 
Further, it can be noticed that the MSMC models perform equally well for all the considered languages, and they are not biased towards any language. The results portray that the Int-2-SMC model has better average classification accuracy across the languages. Hence, this integrated SMC model can be used as a front end of multilingual PRS for optimally recognizing the phonetic units present in the input speech signal of any mode spoken in any language.

\subsection{Performance analysis of integrated model}\label{fine}
 The improved performance of the integrated model (Int-2-SMC) can be better understood from Figure \ref{com}.  
The mode classification using VT-SMC, Src-SMC and Int-2-SMC models are shown in Figure \ref{com} for 25 random utterances from the multilingual test dataset (discussed in Section \ref{corpus}). The selected utterances belong to read mode.
In Figure \ref{com}, the red color circle is representing the correctly detected mode, and the black color circle is representing the falsely detected mode of an utterance using the model shown in the y-axis. 
The performance of Int-2-SMC model depends on the classification accuracy of VT-SMC and Src-SMC models. 
It can be visualized from Figure \ref{com} that if both models fail, then the Int-2-SMC model will also fail in accurately classifying the mode. For example, the mode of utterance 15 is falsely detected in VT-MSC and Src-SMC models as well as in Int-2-SMC model.
Further, analyzed that Src-SMC model recognizes some utterances which are not recognized by VT-SMC model. Similarly, VT-SMC model recognizes some utterances which are not recognized by Src-SMC model.
For example, the mode of utterances 2 and 4 are recognized correctly using Src-SMC model whereas VT-SMC model is not able to recognize it correctly. Similarly, the mode of utterances 5 and 6 are recognized correctly by VT-SMC model, whereas Src-SMC model is not recognizing correctly. However, in the integrated model (Int-2-SMC) mode of utterances 2,4,5 and 6 are recognized correctly. This shows that vocal tract and source features contain some complementary mode-specific information which results in better classification accuracy with Int-2-SMC model.
 \begin{figure}[t]
\centering
\includegraphics[width=0.8\textwidth]{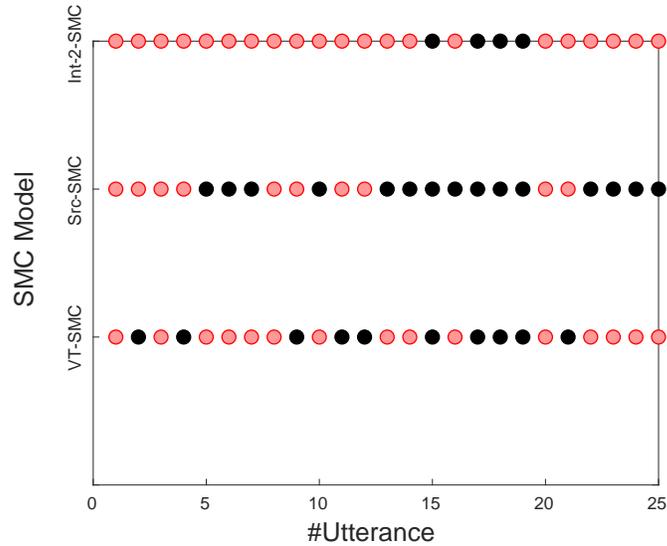}
\caption{Illustration of mode classification for a subset of evaluation utterances using VT-SMC, Src-SMC, and Int-2-SMC models. 
Correctly detected mode of an utterance is marked with the red circle, and falsely detected mode of an utterance is marked with the black circle.\label{com}}
\end{figure}

\subsection{Performance comparison of two speech modes}\label{s14}
In this section, we have analyzed the classification accuracies for conversation and read modes of each monolingual and multilingual speech corpora. 
Here, we have considered the best SMC model (labeled as the Int-2-SMC model) for calculating the performance for individual modes of each language in Table \ref{mode}. 
In the first column of the Table \ref{mode}, mode performances of monolingual models are shown with IDs: (1-4) and for multilingual models are shown with IDs: (5-9).
The second and third columns of Table \ref{mode} depicts the various languages involved in training and testing the models. 
 The fourth and fifth column shows the performance of conversation and read modes speech data. From the results, it can be observed that the classification performance for read mode is better than the conversation speech mode throughout the monolingual and multilingual models. 
 In this work, read speech contains TV news reading style, and the conversation speech contains TV news interview style. In read mode, a speaker utters in very restricted environment, whereas in the conversation mode speakers have no restriction while speaking, and sometimes speech can have read mode characteristics. This could be the reason for the better performance of read mode compared to the conversation mode. 
 It is observed that the interviews mostly starts with a neutral expression. This may be the cause of the misclassification of conversation speech. On the other hand, speakers in read mode are emphasizing some keywords which result in significant variation in acoustic characteristics. This may be the reason for misclassification of read speech.
 
 The pitch contour is considered for better understanding the acoustic and linguistic variations among the read and conversation modes. Informally, we have analyzed the pitch variation on 30 utterances for each mode of a language. It is observed that pitch pattern is varying differently in each mode of a language; however, this variation is similar to the specific mode of a language irrespective of the speakers. For visualizing this analysis, examples of pitch contour are plotted for read and conversation modes using 16 distinct utterances spoken by 8 distinct (4-male and 4-female) speakers in Figure \ref{pit}. Here, 2 sentences are selected from each speaker for plotting pitch contour. Figures 1(a), and 1(b) represents the pitch contour of 8 distinct Bengali utterances in read and conversation modes, respectively. However, Figures 1(c), and 1(d), shows the pitch contour of 8 distinct Odia utterances in read and conversation modes, respectively. It can be observed from Figures 1(a,c) and 1(b,d) that the pitch contour corresponding to conversation mode is more dynamic compared to read mode, irrespective of language and speaker. The presence of expression in conversation mode causes significant pitch variation in speech. However, the absence of emotion in read speech results in almost flat pitch contour. As well as, some of the pitch contours in Figure 1(b,d) are behaving like read speech because of less expression in those sentences. Similarly, some of the pitch contours in Figure 1(a,c) are behaving like conversation speech because of highlighting keywords in an utterance. Hence, most of the read and conversation modes are acoustically and linguistically distinct, except for some cases (discussed above) where these modes have similar acoustic characteristics.
 
 From the Table \ref{mode}, it can be observed that the multilingual SMC model (labeled as K-T-B-O) is performing similar to the monolingual SMC model (labeled as Bengali, Odia, Telugu, and Kannada)  for classifying speech modes. 
Hence, the proposed multilingual SMC model can be significantly used for determining mode of input speech spoken in Indian languages which is used for developing the model.

\begin{table}[h]
\begin{center}
\caption{{Classification accuracy (\%) of conversation, and read modes for  monolingual and multilingual based Int-2 SMC models.}\label{mode} }
\scalebox{1}{
\begin{tabular}{|c|cc|cc|}
\hline
\multirow{2}{*}{\textbf{ID}}&\multicolumn{2}{c|}{\bf{Language}} & \multicolumn{2}{c|}{\bf{Accuracy (\%)}}  \\ \cline{2-5}
&\textbf{Training} & \textbf{Testing} & \textbf{Conversation} & \textbf{Read}\\ 
\hline
1&{Bengali (B)}& Bengali & 92.47&93.85 \\

2&{Odia (O)}& Odia & 93.37& 95.20 \\

3&{Telugu (T)}& Telugu & 89.76& 93.58\\

4&{Kannada (K)}& Kannada & 89.10&92.63 \\\hline

5&\multirow{4}{*}{K-T-B-O}& Bengali & 91.42 & 92.48 \\
6&& Odia & 91.35 & 93.83\\
7&& Telugu & 89.32 & 91.76\\
8&& Kannada & 87.81 & 91.05\\
9&& \textbf{Average} & 89.97 & 92.28\\
\hline
\end{tabular}}
\end{center}
\end{table}

\section{Evaluation of phone recognition systems}\label{eva_mprs}
 In this section, we have evaluated the performance of the baseline and the proposed MPRSs. The baseline and proposed MPRSs are trained and tested using data from the four Indian languages: Telugu, Kannada, Odia, and Bengali discussed in Section \ref{corpus}. In this work, two separate baseline MPRSs are developed one for each read and conversation modes and named as read-baseline and conv-baseline MPRSs, respectively. The baseline mode-dependent MPRSs are developed using MFCCs and tandem features. On the other hand, two separate proposed MPRSs are developed one for each read and conversation modes and named as read-proposed, and conv-proposed MPRSs. A two-stage system is proposed named as a COMB-MPRS, this includes a speech mode classifier (with maximum performance of 91.10\% as shown in Table \ref{msmc_per}) at the first stage, which acts as a switch between the mode-dependent MPRSs (read-proposed MPRS and conv-proposed MPRS) present in the second stage (see Figure \ref{mprs}). The proposed mode-dependent MPRSs is developed using a combination of MFCCs, tandem, and two excitation source features: MPDSSs and RMFCCs. The detailed description of the development of the SMC model is discussed in Section \ref{prop_msmc}.

For performance evaluation, the decoded transcription is matched with the original transcription, and the given Equation (\ref{6}) is used for computing the Phone Error Rate ($E$).

\begin{equation}\label{6}
E=\frac{S+D+I}{N}
\end{equation}

Where $N$ represents the total number of phones in the
original transcription, $D$ accounts for the number of deletion errors, $S$ represents the number of substitution errors and $I$ accounts for the number of insertion errors in the decoded output. 

\begin{table}[h]
\begin{center}
\caption{Recognition accuracy (\%) of phones from read and conversation (Conv) modes of speech in the presence of baseline and proposed MPRSs.\label{table_mprs} }
\scalebox{1.04}{
\begin{tabular}{|c|c|ccc|}
\hline
\multirow{2}{*}{\bf{Feature}}  &\multirow{2}{*}{\bf{MPRS}}  & \multicolumn{3}{c|}{\bf{Accuracy (\%)}} \\ \cline{3-5}
 & & \bf{Read} & \bf{Conversation}& \bf{Average}\\ 
\hline
\multirow{1}{*}{{MFCC+Tandem}} 
&{Read-Baseline}& 64.45&33.50 & 48.97\\
\multirow{1}{*}{{(Baseline)}} 
&{Conv-Baseline}& 30.30&64.23 & 47.26\\\hline
\multirow{2}{*}{{MFCC+Tandem+}}  
&{Read-proposed}& 68.15&38.31 & 53.23\\
\multirow{2}{*}{RMFCC+MPDSS}
&{Conv-proposed}& 34.85&66.53 & 50.69\\
&{COMB-MPRS }& 61.02&59.37 & \textbf{60.19}\\
\hline
\end{tabular}}
\end{center}
\end{table}
 Table \ref{table_mprs} depicts the phone recognition accuracies of baseline and proposed systems. The first column of Table \ref{table_mprs} represent the various features used for developing the MPRSs. The second column shows the name of the recognition systems. In this work, the recognition accuracy of each system is computed at a detailed and finer level. When a system is tested without the knowledge of speech mode, it will provide detailed analysis which is shown in column fifth. When a system is tested separately for read and conversation mode speech, it will provide finer level information which is shown in column third and fourth. 
 The first and third rows contain the phone recognition accuracies for read-baseline and read-proposed MPRSs, which are trained with read mode data. In the second and fourth rows, the performance of conv-baseline and conv-proposed MPRSs are shown, which are trained using conversation mode of speech corpora. The fifth row is depicting the accuracy of the COMB-MPRS, which includes read-proposed and conv-proposed MPRSs, therefore, read, and conversation speech corpora are used for training the mode-specific proposed MPRSs.
  

From Table \ref{table_mprs}, it is observed that the average performance of the proposed MPRSs is better than the baseline MPRSs. The addition of excitation source features (RMFCC and MPDSS) has contributed to improvement in phone recognition accuracies with proposed MPRSs. This could be analyzed by comparing the performance of the mode-dependent proposed MPRSs with the mode-dependent baseline MPRSs. An overall improvement of 4.26\% and 3.43\% are achieved with the read-proposed MPRS compared to read-baseline MPRS and the conv-proposed MPRS compared to conv-baseline MPRS, respectively. This shows that the excitation source features contain distinct phone related information. 

 In Table \ref{table_mprs}, the average accuracies for read-proposed and conv-proposed MPRSs are depicting that the testing data processed through the systems without the knowledge of the mode of input speech. For analyzing the accuracy at the finer level, we have computed the recognition accuracy for each mode separately using the read-proposed and conv-proposed MPRSs. Since the prior knowledge about the modes of the testing data is given, thus, we have recognized the accuracy at a finer level. It can be observed that the read-proposed MPRS is providing a significantly better result for read mode as compared to conv-proposed MPRS. For the conversation mode, the significant result is provided by conv-proposed MPRS than the read-proposed MPRS. Hence, the best performance for read and conversation modes that could be achieved using the COMB-MPRS in an ideal case (100\% MSMC accuracy) are 68.15\% and 66.53\%, respectively. Therefore, the average of best accuracies achieved for read and conversation modes will represent the overall accuracy that could be attained by the COMB-MPRS in an ideal case, which is 67.34\%.
However, the achieved performance for the COMB-MPRS having an MSMC model with maximum performance (91.10\%) is 60.19\%, which is 7.15\% smaller than the ideal case. This is because the MSMC model is not accurately classifying the modes of some utterances. 
But, on an average, the COMB-MPRS is performing much better than the mode-dependent baseline and proposed MPRSs. The overall improvement of 11.22\%, 12.93\%, 6.96\%, and 9.5\% are achieved with the COMB-MPRS compared to read-baseline, conv-baseline, read-proposed, and conv-proposed MPRSs, respectively. The reason is that the input speech in COMB-MPRS is processed through the corresponding mode-specific MPRS. Here, the mode of the speech utterance is identified using the speech mode classification model placed at the front end.
This shows the importance of the MSMC. Therefore, incorporating MSMC is important, and also in the future, we have to make it more closer to ideal accuracy.
In Table \ref{table_mprs}, it is analyzed that the recognition accuracy achieved using the proposed system for read mode (61.02\%) is better than the conversation mode phones (59.37\%). This is because the multilingual SMC model at the front end of MPRS is giving better classification performance for read mode (92.28\%) than the conversation mode (89.97\%) as shown in Table \ref{msmc_per}. The overall result indicates that combining mode-specific MPRSs and MSMC into a single framework can better decode the phonetic units present in a speech utterance of any language spoken in any mode.

\section{Conclusion}\label{con}
In this work, a system has been proposed for automatically recognizing phonetic units present in speech utterances from multiple languages spoken in multiple modes. In this study, the considered modes of speech are conversation and read modes in four Indian languages, namely, Telugu, Kannada, Odia, and Bengali. The proposed method is explored at two-stage.
In the first stage, the mode of input speech is identified using a multilingual SMC system (K-T-B-O) developed using vocal-tract and suprasegmental level excitation source features. The vocal-tract information provides better SMC accuracy (82\%) compared to excitation source information (69\%). Experimental analysis indicates that there exists distinct mode-specific information between vocal-tract and excitation source features. {Hence, the scores of excitation source based models are further combined with the scores obtained from vocal tract based model to improve the SMC accuracy (91\%).} In the second stage, speech utterance will be routed to the MPRS of the mode identified in the previous stage, and the phonetic units present in the speech signal are determined. 
The evaluation of phone recognition shows that the proposed two-stage system significantly outperforms the baseline mode-dependent MPRSs. In the future, we intend to investigate other speech features to improve MSMC accuracy as it is required to have an ideal MSMC system with 100\% SMC accuracy in the first-stage. We will also explore articulatory features with vocal tract details for further improving the performance of MPRS. In this work, we have focussed on Indian languages; however, in further study, other languages will be explored for analyzing the significance of the proposed system.


\section*{References}

\bibliography{111}

\begin{thebibliography}{10}
\expandafter\ifx\csname url\endcsname\relax
  \def\url#1{\texttt{#1}}\fi
\expandafter\ifx\csname urlprefix\endcsname\relax\def\urlprefix{URL }\fi
\expandafter\ifx\csname href\endcsname\relax
  \def\href#1#2{#2} \def\path#1{#1}\fi

\bibitem{Manjunath2015}
K.~Manjunath, K.~S. Rao, Source and system features for phone recognition,
  International Journal of Speech Technology 18~(2) (2015) 257--270.
\newblock \href {http://dx.doi.org/10.1007/s10772-014-9266-0}
  {\path{doi:10.1007/s10772-014-9266-0}}.

\bibitem{Pradeep2016}
R.~Pradeep, K.~S. Rao, {Deep neural networks for Kannada phoneme recognition},
  in: Proceedings of Ninth International Conference on Contemporary Computing
  (IC3), JIIT, Noida, 2016, pp. 1--6.
\newblock \href {http://dx.doi.org/10.1109/IC3.2016.7880202}
  {\path{doi:10.1109/IC3.2016.7880202}}.

\bibitem{Scanzio2008}
S.~Scanzio, P.~Laface, L.~Fissore, R.~Gemello, F.~Mana, On the use of a
  multilingual neural network front-end, in: Proceedings of Ninth Annual
  Conference of the International Speech Communication Association, Brisbane,
  Australia, 2008, pp. 2711--2714.

\bibitem{Burget2010}
L.~Burget, P.~Schwarz, M.~Agarwal, P.~Akyazi, K.~Feng, A.~Ghoshal, O.~Glembek,
  N.~Goel, M.~Karafi{\'a}t, D.~Povey, et~al., {Multilingual acoustic modeling
  for speech recognition based on subspace Gaussian mixture models}, in:
  Proceedings of International Conference on Acoustics Speech and Signal
  Processing (ICASSP), Dallas, Texas, 2010, pp. 4334--4337.
\newblock \href {http://dx.doi.org/10.1109/ICASSP.2010.5495646}
  {\path{doi:10.1109/ICASSP.2010.5495646}}.

\bibitem{Manjunath2019}
K.~Manjunath, D.~B. Jayagopi, K.~S. Rao, V.~Ramasubramanian, Development and
  analysis of multilingual phone recognition systems using indian languages,
  International Journal of Speech Technology 22~(1) (2019) 157--168.
\newblock \href {http://dx.doi.org/10.1007/s10772-018-09589-z}
  {\path{doi:10.1007/s10772-018-09589-z}}.

\bibitem{Kumar2005}
C.~S. Kumar, V.~Mohandas, H.~Li, Multilingual speech recognition: A unified
  approach, in: Proceedings of Ninth European Conference on Speech
  Communication and Technology, Lisbon, Portugal, 2005, pp. 3357--3360.

\bibitem{Gangashetty2005}
S.~V. Gangashetty, C.~C. Sekhar, B.~Yegnanarayana, Spotting multilingual
  consonant-vowel units of speech using neural network models, in: Proceedings
  of International Conference on Nonlinear Analyses and Algorithms for Speech
  Processing, Berlin, Heidelberg, 2005, pp. 303--317.
\newblock \href {http://dx.doi.org/10.1007/11613107_27}
  {\path{doi:10.1007/11613107_27}}.

\bibitem{Mohan2014}
A.~Mohan, R.~Rose, S.~H. Ghalehjegh, S.~Umesh, {Acoustic modelling for speech
  recognition in Indian languages in an agricultural commodities task domain},
  Speech Communication 56 (2014) 167--180.
\newblock \href {http://dx.doi.org/10.1016/j.specom.2013.07.005}
  {\path{doi:10.1016/j.specom.2013.07.005}}.

\bibitem{Batliner1995}
A.~Batliner, R.~Kompe, A.~Kie{\ss}ling, E.~N{\"o}th, H.~Niemann, Can you tell
  apart spontaneous and read speech if you just look at prosody?, in: Speech
  Recognition and Coding, Springer, 1995, pp. 321--324.
\newblock \href {http://dx.doi.org/10.1007/978-3-642-57745-1_47}
  {\path{doi:10.1007/978-3-642-57745-1_47}}.

\bibitem{Blaauw1991}
E.~Blaauw, Phonetic characteristics of spontaneous and read-aloud speech, in:
  Phonetics and Phonology of Speaking Styles, 1991.

\bibitem{Dellwo2015}
V.~Dellwo, A.~Leemann, M.-J. Kolly, The recognition of read and spontaneous
  speech in local vernacular: The case of zurich german, Journal of Phonetics
  48 (2015) 13--28.
\newblock \href {http://dx.doi.org/10.1016/j.wocn.2014.10.011}
  {\path{doi:10.1016/j.wocn.2014.10.011}}.

\bibitem{Hansen_1989}
J.~H. Hansen, Analysis and compensation of stressed and noisy speech with
  application to robust automatic recognition, Signal Processing 17~(3) (1989)
  282.
\newblock \href {http://dx.doi.org/10.1016/0165-1684(89)90010-8}
  {\path{doi:10.1016/0165-1684(89)90010-8}}.

\bibitem{Rostolland1982a}
D.~Rostolland, Phonetic structure of shouted voice, Acta Acustica united with
  Acustica 51~(2) (1982) 80--89.

\bibitem{Rostolland1982}
D.~Rostolland, Acoustic features of shouted voice, Acta Acustica united with
  Acustica 50~(2) (1982) 118--125.

\bibitem{Zhang2007}
C.~Zhang, J.~H. Hansen, Analysis and classification of speech mode: whispered
  through shouted, in: Eighth Annual Conference of the International Speech
  Communication Association, Antwerp, Belgium, 2007, pp. 2289--2292.

\bibitem{Dede2010}
G.~Dede, M.~H. Sazl{\i}, Speech recognition with artificial neural networks,
  Digital Signal Processing 20~(3) (2010) 763--768.
\newblock \href {http://dx.doi.org/10.1016/j.dsp.2009.10.004}
  {\path{doi:10.1016/j.dsp.2009.10.004}}.

\bibitem{Vinyals2011}
O.~Vinyals, S.~V. Ravuri, Comparing multilayer perceptron to deep belief
  network tandem features for robust asr, in: international conference on
  acoustics, speech and signal processing (ICASSP), IEEE, 2011, pp. 4596--4599.
\newblock \href {http://dx.doi.org/10.1109/icassp.2011.5947378}
  {\path{doi:10.1109/icassp.2011.5947378}}.

\bibitem{Koolagudi2012}
S.~G. Koolagudi, K.~S. Rao, Emotion recognition from speech: a review,
  International journal of speech technology 15~(2) (2012) 99--117.
\newblock \href {http://dx.doi.org/10.1007/s10772-011-9125-1}
  {\path{doi:10.1007/s10772-011-9125-1}}.

\bibitem{Kumar2013}
S.~S. Kumar, K.~S. Rao, D.~Pati, {Phonetic and prosodically rich transcribed
  speech corpus in Indian languages: Bengali and Odia}, in: Proceedings of
  International Conference on Oriental COCOSDA held jointly with Conference on
  Asian Spoken Language Research and Evaluation (O-COCOSDA/CASLRE), Gurgaon,
  India, 2013, pp. 1--5.
\newblock \href {http://dx.doi.org/10.1109/icsda.2013.6709901}
  {\path{doi:10.1109/icsda.2013.6709901}}.

\bibitem{Shridhara2013}
M.~Shridhara, B.~K. Banahatti, L.~Narthan, V.~Karjigi, R.~Kumaraswamy,
  {Development of Kannada speech corpus for prosodically guided phonetic search
  engine}, in: Proceedings of international conference on oriental COCOSDA held
  jointly with conference on Asian spoken language research and evaluation
  (O-COCOSDA/CASLRE), Gurgaon, India, 2013, pp. 1--6.
\newblock \href {http://dx.doi.org/10.1109/icsda.2013.6709875}
  {\path{doi:10.1109/icsda.2013.6709875}}.

\bibitem{Murty2008}
K.~S.~R. Murty, B.~Yegnanarayana, Epoch extraction from speech signals, IEEE
  Transactions on Audio, Speech, and Language Processing 16~(8) (2008)
  1602--1613.
\newblock \href {http://dx.doi.org/10.1109/tasl.2008.2004526}
  {\path{doi:10.1109/tasl.2008.2004526}}.

\bibitem{Yegnanarayana2009}
B.~Yegnanarayana, K.~S.~R. Murty, Event-based instantaneous fundamental
  frequency estimation from speech signals, IEEE Transactions on Audio, Speech,
  and Language Processing 17~(4) (2009) 614--624.
\newblock \href {http://dx.doi.org/10.1109/tasl.2008.2012194}
  {\path{doi:10.1109/tasl.2008.2012194}}.

\bibitem{Lam1997}
L.~Lam, S.~Suen, Application of majority voting to pattern recognition: an
  analysis of its behavior and performance, IEEE Transactions on Systems, Man,
  and Cybernetics-Part A: Systems and Humans 27~(5) (1997) 553--568.
\newblock \href {http://dx.doi.org/10.1109/3468.618255}
  {\path{doi:10.1109/3468.618255}}.

\bibitem{Benesty2009}
J.~Benesty, J.~Chen, Y.~Huang, I.~Cohen, Pearson correlation coefficient, in:
  Noise reduction in speech processing, Springer, 2009, pp. 1--4.
\newblock \href {http://dx.doi.org/10.4135/9781412953948.n342}
  {\path{doi:10.4135/9781412953948.n342}}.

\bibitem{Povey2011}
D.~Povey, A.~Ghoshal, G.~Boulianne, L.~Burget, O.~Glembek, N.~Goel,
  M.~Hannemann, P.~Motlicek, Y.~Qian, P.~Schwarz, et~al., {The Kaldi speech
  recognition toolkit}, in: workshop on automatic speech recognition and
  understanding, no. EPFL-CONF-192584, IEEE Signal Processing Society, 2011.

\bibitem{Zhang2014}
X.~Zhang, J.~Trmal, D.~Povey, S.~Khudanpur, Improving deep neural network
  acoustic models using generalized maxout networks, in: International
  Conference on Acoustics, Speech and Signal Processing (ICASSP), IEEE,
  Florence, Italy, 2014, pp. 215--219.
\newblock \href {http://dx.doi.org/10.1109/icassp.2014.6853589}
  {\path{doi:10.1109/icassp.2014.6853589}}.

\bibitem{Yu2015}
D.~Yu, L.~Deng, Deep neural network-hidden markov model hybrid systems, in:
  Automatic Speech Recognition, Springer, 2015, pp. 99--116.
\newblock \href {http://dx.doi.org/10.1007/978-1-4471-5779-3_6}
  {\path{doi:10.1007/978-1-4471-5779-3_6}}.

\bibitem{Brown1992}
P.~F. Brown, P.~V. Desouza, R.~L. Mercer, V.~J.~D. Pietra, J.~C. Lai,
  Class-based n-gram models of natural language, Computational linguistics
  18~(4) (1992) 467--479.

\bibitem{Svozil1997}
D.~Svozil, V.~Kvasnicka, J.~Pospichal, Introduction to multi-layer feed-forward
  neural networks, Chemometrics and intelligent laboratory systems 39~(1)
  (1997) 43--62.
\newblock \href {http://dx.doi.org/10.1016/S0169-7439(97)00061-0}
  {\path{doi:10.1016/S0169-7439(97)00061-0}}.

\bibitem{Csaji2001}
B.~C. Cs{\'a}ji, Approximation with artificial neural networks, Faculty of
  Sciences, Etvs Lornd University, Hungary 24 (2001) 48.

\bibitem{Bottou2010}
L.~Bottou, Large-scale machine learning with stochastic gradient descent, in:
  Proceedings of COMPSTAT'2010, Springer, 2010, pp. 177--186.
\newblock \href {http://dx.doi.org/10.1007/978-3-7908-2604-3_16}
  {\path{doi:10.1007/978-3-7908-2604-3_16}}.

\bibitem{Reddy2013}
V.~R. Reddy, S.~Maity, K.~S. Rao, {Identification of Indian languages using
  multi-level spectral and prosodic features}, International Journal of Speech
  Technology 16~(4) (2013) 489--511.
\newblock \href {http://dx.doi.org/10.1007/s10772-013-9198-0}
  {\path{doi:10.1007/s10772-013-9198-0}}.

\end{thebibliography}

\end{document}